\documentclass[]{article}

\usepackage{authblk}
\usepackage{textcomp}
\usepackage[T1]{fontenc}
\usepackage[bitstream-charter]{mathdesign}
\usepackage[symbol]{footmisc}

\usepackage{fancyhdr}
\usepackage{amsmath}
\usepackage{bm}
\usepackage{booktabs}
\usepackage{color}
\usepackage{graphicx}
\usepackage{caption}
\usepackage{siunitx}
\usepackage[UKenglish]{isodate}
\usepackage{eso-pic,graphicx}
\usepackage[top=1.2in, bottom=1.2in, outer=1.2in, inner=1.2in]{geometry}
\usepackage{lastpage}
\usepackage{multicol}
\usepackage[version=4]{mhchem}
\usepackage{titlesec}
\usepackage{subcaption}
\usepackage{epstopdf}
\usepackage{parboxx}

\usepackage{verbatim}
\usepackage{listing,lstbayes}
\usepackage{cite}
\usepackage{enumitem}

\usepackage{setspace}
\usepackage{lineno}
\usepackage{todonotes}

\definecolor{mygreen}{rgb}{0,0.6,0}
\definecolor{mygray}{rgb}{0.5,0.5,0.5}
\definecolor{mymauve}{rgb}{0.58,0,0.82}

\lstdefinestyle{interfaces}{
	float,
	floatplacement=tbp
}

\DeclareCaptionFormat{listing}{#1#2#3}
\title{Revision of ISO~19229 to support the certification of calibration gases for purity\footnotemark{}\footnote{Submitted to Accreditation and Quality Assurance}}
\author{Adriaan M.H. van der Veen} 
\author{Gerard Nieuwenkamp}
\affil{VSL, Thijsseweg 11, 2629 JA Delft, the Netherlands}

\cleanlookdateon

\begin{document}

\maketitle

\begin{abstract}
The second edition of ISO~19229 expands the guidance in its predecessor in two ways. Firstly, it provides more support and examples describing possible experimental approaches for purity analysis. A novelty is that it describes how the beta distribution, or some other suitable probability distribution can be used to approximate the distribution of the output quantity, i.e., the fraction of a component. It also provides guidance on how to report coverage intervals in those cases, where the usual approximation from the GUM (Guide to the expression of Uncertainty in Measurement) to use the normal or $ t $ distribution is inappropriate because of vicinity of zero. Coverage intervals play an important role in conformity assessment, and it is also customary to report measurement uncertainty in the form of a coverage interval, notwithstanding that ISO/IEC~17025 does not explicitly require it. ISO~6141, which sets requirements for certificates of calibration gas mixtures, does require the statement of an expanded uncertainty, which has been interpreted that in the case of a non-symmetric output probability distribution, a coverage interval should be stated, along with the value and the standard uncertainty. This paper gives a brief background to the choices made and examples in ISO~19229.
\end{abstract}

\section*{Key words}

beta distribution; normal distribution; purity analysis; coverage interval; ISO~19229.

\section{Introduction}

ISO~19229 \cite{ISO19229} was originally developed to support documentary standards describing static \cite{ISO6142a,ISO6144} and dynamic methods \cite{ISO6145} for calibration gas mixture preparation. As the latter documentary standards are based on the law of propagation of uncertainty from the Guide to the expression of Uncertainty in Measurement (GUM) \cite{GUM}, it sufficed to provide guidance for calculating values and standard uncertainties for the fractions of impurities. For the provision of calibration and reference material certificates, the guidance fell short in that it is not always appropriate to base an expanded uncertainty of a fraction on the normal or $ t $ distribution. 

The first edition of ISO~19229 \cite{ISO19229} has been revised to provide more guidance and examples on the experimental aspects of purity analysis, also related to applications of ``zero gas'' (the gas used to establish a zero point for an analyser). Such applications are often related to the calibration of an analyser \cite{ISO6143,ISO12963}. The second edition of ISO~19229 \cite{ISO19229:2018} also considers the idea that the producer of the ``zero gas'' is not the user, so that a certificate has to be issued. Although it is not explicitly stated in ISO/IEC~17025 \cite{ISO17025}, measurement uncertainty is usually reported as an expanded uncertainty, from which the recipient can derive the standard uncertainty using the stated coverage factor. ISO~6141 \cite{ISO6141} is more explicit and requires the provision of an expanded uncertainty, which is the half-width of a (symmetric) coverage interval \cite{GUM}.

As the results of purity analysis are often close to zero (for a fraction of an impurity) or one (for the fraction of the most abundant component), it is often inappropriate to approximate the probability density function of the output quantity by the normal distribution or $ t $ distribution, as presumed by the GUM \cite{GUM}. Instead, often a non-symmetric probability density function is required, to avoid that part of the coverage interval lies in an infeasible region (below zero or above one for fractions). 
Specific guidance is provided in the second edition of ISO~19229 to establish coverage intervals that are valid and can be used in applications that require coverage intervals, such as assessing compliance with a limit \cite{JCGM106}, the development of specifications, and the comparison of measurement results. 

In this paper, some background is given as to the choices made in the development of the clause on establishing coverage intervals, and some more elaborate explanations. 

\section{Coverage intervals}

The most commonly used probability density functions in metrology are the normal distribution, $ t $ distribution and rectangular distribution \cite{GUM,GUM-S1}. They are frequently used in type~B evaluations of standard uncertainty. Particularly, the normal distribution and $ t $-distribution are also commonly used to approximate the probability density function of the output quantity (the measurand) \cite{GUM}. In many cases, such an approximation is adequate for generating \SI{95}{\percent} coverage intervals. Since the publication of the Monte Carlo method of GUM Supplement 1 (GUM-S1) \cite{GUM-S1}, there is a more reliable way of obtaining a coverage interval, namely by propagating the assigned probability density functions to the input quantities using the measurement model. This method has not been chosen for the second edition of ISO~19229, for most practitioners rely on the law of propagation of uncertainty from the GUM \cite{GUM} to propagate the uncertainties arising during purity analysis. 

As the Monte Carlo method of GUM-S1 will usually provide a non-symmetric coverage interval, the reporting requirements have been articulated as follows \cite{GUM-S1}:
\begin{enumerate} [label=\alph*),noitemsep]
	\item an estimate $ y $ of the output quantity $ Y $; 
	\item the standard uncertainty $ u(y) $ associated with $ y $;
	\item the stipulated coverage probability $ 100p $~\si{\percent} (e.g. \SI{95}{\percent}); 
	\item the endpoints of the selected $ 100p $~\si{\percent} coverage interval (e.g. \SI{95}{\percent} coverage interval) for $ Y $; 
	\item any other relevant information, such as whether the coverage interval is a probabilistically symmetric coverage interval or a shortest coverage interval
\end{enumerate}
It is important to note that for non-symmetric coverage intervals, there is 
\begin{enumerate} [noitemsep]
	\item a difference between a probabilistically symmetric coverage interval and a shortest coverage interval. For symmetric probability density functions, these intervals are the same;
	\item no way to calculate from a non-symmetric coverage interval the standard uncertainty;
	\item no expanded uncertainty, neither a coverage factor.
\end{enumerate}
Sometimes a non-symmetric coverage interval is denoted by introducing $ U_- $ and $ U_+ $, where $ U_- $ denotes the width of the coverage interval from its lower bound to the estimate (= measured value) and $ U_+ $ denotes the width of the coverage interval from the estimate to its upper bound. 

ISO~19229 \cite{ISO19229:2018} does not specify how the coverage interval is reported if it is symmetric. If the interval is non-symmetric, ISO~19229 \cite{ISO19229:2018} requires that the bounds of it shall be reported, as well as the estimate and standard uncertainty, which are required for using the law of propagation of uncertainty from the GUM \cite{GUM} in a subsequent application. These reporting requirements appear at first glance more involved than those usually applied (an expanded uncertainty, a probability level and a coverage factor), but these follow largely good practice in reporting non-symmetric coverage intervals as described in GUM-S1 \cite{GUM-S1}. The various elements of the reporting focus (potentially) on different users with different requirements. 

\section{Purity analysis}

The result of a purity analysis of a component $ i $ in a ``pure'' material, when expressed as a fraction, is by definition a non-negative value $ x_i $ and associated standard uncertainty $ u(x_i) $. The standard uncertainty in chemical composition measurement is often expressed as a relative standard uncertainty, i.e., $ u(x_i)/x_i $, which is admissible as long as $ x_i > 0 $. It should be recalled that the standard uncertainty is not only a matter of repeatability of measurement; effects such as the uncertainty associated with the calibration, the measurement of the ``zero'' (blank) etc. may lead to large values for the relative standard uncertainty for small values of $ x_i $. ISO~19229 \cite{ISO19229:2018} considers the estimate $ x_i $ as close to zero if $ x_i \leq 4 u(x_i) $, bearing in mind that usually a \SI{95}{\percent} coverage interval is to be reported. A different value for the factor than 4 may be chosen for other coverage probabilities. 

Using the normal distribution and a coverage probability $ p = 0.95 $, the coverage factor $ k = 1.96 $ \cite{GUM}. The coverage interval for this probability density function is symmetrical, so it can given as $ x_i \pm U(x_i)$, where $ U(x_i) $ denotes the expanded uncertainty associated with $ x_i $. The coverage interval thus obtained is the shortest, leaving $ 0.025 $ of the probability below the lower bound $ x_i - U(x_i) $ and $ 0.025 $ above the upper bound $ x_i + U(x_i) $. 

In the vicinity of zero, the lower bound of the normal coverage interval $ x_i - U(x_i) $ may drop below zero. In these cases, the normal distribution is no longer acceptable to establish a coverage interval. A value is close to zero if the number of standard uncertainties between it and 0 is small (say, less than 4). So, for example if $ x_i = \SI{3.0e-9}{\mole\per\mole} $ with $ u(x_i) = \SI{2.0e-9}{\mole\per\mole} $, then $ x_i $ is in the vicinity of zero.  For obtaining a coverage interval, a probability density function is required that is defined on the interval only, preferably between 0 and 1, the interval on which the amount-of-substance fraction is defined. 

\begin{figure}
	\centering
	\includegraphics[width=0.95\textwidth]{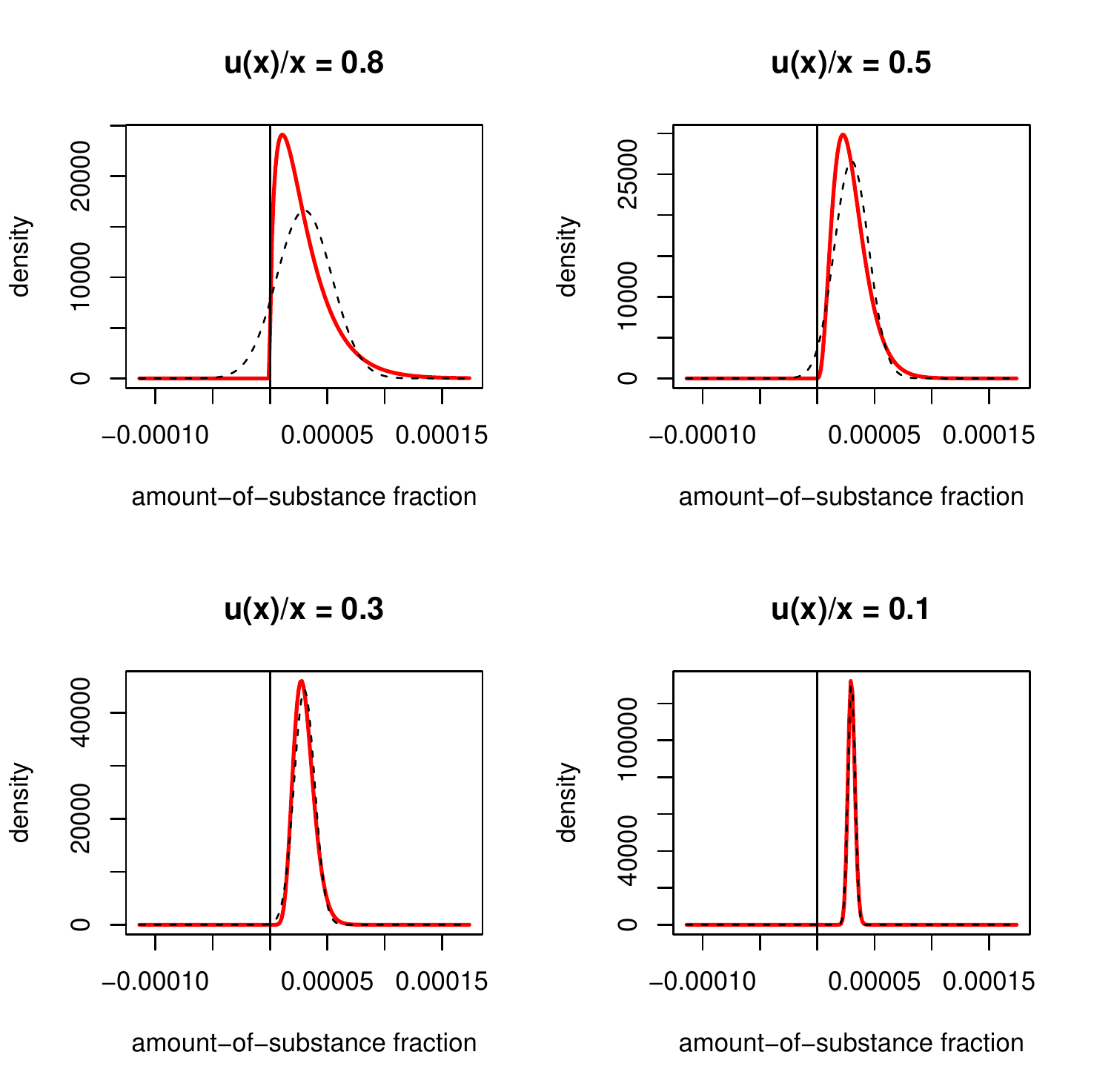} 
	\caption{Comparison between the normal (black dotted line) and beta distribution (red solid line) for different values of the coefficient of variation. The vertical line indicates zero on the amount-of-substance fraction axis} \label{fig:beta1}
\end{figure}

If the normal distribution is considered to be a reasonable approximation for the probability density function of the amount-of-substance fraction of an impurity which is not close to zero, then one would like to have a probability density function that (1) has a similar shape as the normal distribution under these considerations and (2)  would restrict the amount-of-substance fraction to be between $ [0, 1] $, boundaries included. Such a candidate is the beta distribution, \cite{Gelman3rd}. Figure~\ref{fig:beta1} shows for four different relative standard uncertainty levels the beta distribution and the normal distribution. According to the criterion in ISO~19229 \cite{ISO19229:2018}, only the case where $ u(x)/x = 0.1 $ is considered to be not in the vicinity of zero, the other cases are. It is seen that for a relative standard uncertainty of \SI{10}{\percent} (bottom-right panel in figure~\ref{fig:beta1}), the beta distribution and normal distribution look very similar, and they become less similar with increasing relative standard uncertainty. It is also seen that the area under the curve of the normal probability density function that lies below zero increases with increasing relative standard uncertainty. 

\section{Coverage intervals from the beta distribution}

The computations with the beta distribution are somewhat more involved than with the normal distribution, which is characterised by the mean $ \mu $ and variance $ \sigma^2 $ as parameters. The probability density function of the beta distribution also has two parameters, $ \alpha $ and $ \beta $. The mean and variance of the beta distribution are given by respectively
\begin{equation} \label{eq:mu_beta}
\mu = \dfrac{\alpha}{\alpha + \beta}
\end{equation}
and
\begin{equation} \label{eq:s2_beta}
\sigma^2 = \dfrac{\alpha\beta}{(\alpha + \beta)^2 (\alpha + \beta + 1)}
\end{equation}
Based on equations \eqref{eq:mu_beta} and \eqref{eq:s2_beta}, the parameters of the beta distribution can be expressed in terms of the mean $ \mu $ and variance $ \sigma^2 $, i.e.,
\begin{equation} \label{eq:alpha}
\alpha = \left(\dfrac{1-\mu}{\sigma^2} - \dfrac{1}{\mu}\right) \mu^2
\end{equation}
and
\begin{equation} \label{eq:beta}
\beta = \alpha \left(\dfrac{1}{\mu} -1\right)
\end{equation}
Substituting the values $ x_i $ for $ \mu $ and $ u(x_i) $ for $ \sigma $ enables the calculation of the parameters of the beta distribution from the measured amount-of-substance fraction and its standard uncertainty. Then, with the parameters of the beta distribution, the coverage interval can be calculated. 

For calculating the coverage interval, a probability needs to be specified, say $ p = 0.95 $. Secondly, a decision needs be taken on how to distribute the remaining probability $ 1-p $. Distributing this probability evenly (as is done with the symmetric normal interval above) no longer gives the shortest coverage interval if the distribution is not symmetric. In principle, it is not required to specify the shortest interval, but in particular for symmetrical distribution functions it is customary to do so.  A comparison between the probabilistically symmetric coverage intervals computed from the two distributions is shown in table~\ref{tab:CI_comp} for the same mean and standard deviation. 

\begin{table}[htbp]
  \centering
  \caption{Comparison between coverage intervals from the normal and beta distributions. L denotes the lower bound, H the upper bound of the interval. All data are given as amount-of-substance fractions ($ \times \num{E9} $)}
    \begin{tabular}{S[table-format=3.0]S[table-format=3.0]S[table-format=3.0]S[table-format=3.0]S[table-format=3.0]S[table-format=3.0]}
    \addlinespace
    \toprule
    & & \multicolumn{2}{c}{normal} & \multicolumn{2}{c}{beta} \\
    \cmidrule(lr{1em}){3-4}
    \cmidrule(lr{1em}){5-6}
    {$ x $}     & {$ u(x) $}  & {L}     & {H}     & {L}     & {H} \\
    \midrule
		300   & 240   & -170   & 770   & 23    & 920 \\
		300   & 150   & 6    & 594   & 82   & 657 \\
		300   & 90    & 124   & 476   & 150   & 500 \\
		300   & 30    & 241   & 359   & 244   & 362 \\
    \bottomrule
    \end{tabular}%
  \label{tab:CI_comp}%
\end{table}%

Table~\ref{tab:CI_comp} shows the coverage intervals using the normal distribution ($ k = 1.96 $) and the beta distribution for an amount-of-substance fraction $ x = \SI{300}{\nano\mole\per\mole} $ and four different relative standard uncertainties: \SI{80}{\percent}, \SI{50}{\percent}, \SI{30}{\percent}, and \SI{10}{\percent}. The corresponding probability density functions are shown in figure~\ref{fig:beta1}. It is readily seen that with decreasing standard uncertainty, the lower and upper limits of both coverage intervals become more alike. For a relative standard uncertainty of \SI{10}{\percent} (the last line in table~\ref{tab:CI_comp}), the differences are insignificant. Even for a relative standard uncertainty of \SI{30}{\percent}, the differences are in most practical circumstances meaningless. The calculations shown here provide a rationale for considering a result with a relative standard uncertainty of \SI{25}{\percent} as being ``close to zero'' \cite{ISO19229:2018}. 

\section{Shortest coverage interval}

Instead of a probabilistically symmetric coverage interval, the shortest coverage interval may be required. For a symmetric probability distribution, these intervals coincide \cite{GUM-S1}, but for a non-symmetric coverage interval, they can be markedly different. The shortest interval can be found by ``trial and error'', starting with the probabilistically-symmetric coverage interval. By changing the lower limit (or upper limit), it can be seen whether the width increases or decreases. For some value of this limit, the width of the interval reaches a minimum. A more sophisticated approach is to use a minimisation routine, such as the function \texttt{optim()} in \texttt{R} \cite{R2017}, which implements, among others, Brent's bisection method \cite{NR}. The code of this optimisation is given in section~\ref{sec:ShortestCI}, which is based on the code for computing a coverage interval using the beta distribution (see section~\ref{sec:ShortestCI}). 

Using the example from~\ref{tab:CI_comp}, for a relative standard uncertainty of \SI{30}{\percent}, the shortest coverage interval is given by $ [136, 479] \times \num{E-9} $. The width of this interval is \num{342e-09}, whereas that of the probabilistically symmetric interval is \num{350e-9} (see also table~\ref{tab:CI_comp}). For larger relative standard uncertainties, the difference between the shortest and probabilistically symmetric coverage intervals will increase, depending on the purpose of the coverage interval, the related conformity assessment and what interval is more appropriate \cite{JCGM106}. 

\section{Other probability levels}

It is important to bear in mind that for probabilities other than \SI{95}{\percent}, ``close to zero'' can require a different interpretation. For a \SI{99}{\percent} coverage interval and a relative standard uncertainty of \SI{30}{\percent}, the interval using the normal distribution is $ [68, 532] \times \num{E-9} $ whereas that for the beta distribution the interval is $ [119, 582] \times \num{E-9} $. These intervals are much more different from one another than those given in table~\ref{tab:CI_comp} for \SI{95}{\percent} probability and \SI{30}{\percent} relative standard uncertainty. 


\section{Use of the scaled beta distribution}

In ISO~19229, an example is given where the measured amount-of-substance fraction is \SI{100e-9}{\mole\per\mole} and a standard uncertainty of \SI{30e-9}{\mole\per\mole}. In this example, a scaled beta distribution is used, because the implementation in software of the calculation of the inverse of the cumulative beta distribution is not always sufficiently numerically stable. In older spreadsheet software, it occurred that by directly substituting this measured value and standard uncertainty in equations~\eqref{eq:alpha} and \eqref{eq:beta} and using these parameters in the subsequent calculation resulted in an error message. Newer versions of the same software allowed the direct calculation. 

The calculations with and without scaling are compared as follows. By substituting directly the measured value for $ \mu $ and the squared standard uncertainty for $ \sigma^2 $, $ \mu = \num{100e-9} $ and $ \sigma^2 = \num{9.00E-16} $ in to equation~\eqref{eq:alpha}, $ \alpha = \num{11.11} $. From equation~\eqref{eq:beta}, we obtain $ \beta = \num{1.1111E+08} $. With these values for the coefficients, the \SI{95}{\percent} coverage interval obtained from the inverse beta function is $ [50, 167] \times \num{E-9} $. 

With rescaling of the results by a factor \num{1000}, we obtain $ \mu = \num{100e-6} $ and $ \sigma^2 = \num{9.00E-10} $. From equation~\eqref{eq:alpha}, $ \alpha = \num{11.1099} $ and from~\eqref{eq:beta}, $ \beta = \num{111088} $. With these values for the coefficients, the scaled \SI{95}{\percent} coverage interval obtained from the inverse beta function is $ [50, 167] \times \num{E-6} $. Dividing by the scaling factor gives the coverage interval $ [50, 167] \times \num{E-9} $, as in the case without scaling. When reproducing the limits of the coverage interval with a larger number of digits, it can be shown that the intervals are close, but not identical. The scaled coverage interval is $ [50123, 166809] \times\num{E-9} $, whereas the coverage interval without scaling is $ [50.124, 166.811] \times\num{E-9}$. Apart from the effect of the scaling factor, it is seen that there are tiny differences in the computed limits of the probabilistically symmetric coverage intervals. These differences are however, for representing the result with an adequate number of meaningful digits, irrelevant. 

\section{Perspective of the recipient of the certificate}

Most recipients of calibration or reference material certificates are accustomed to taking the expanded uncertainty $ U $ and use this parameter for a further application, which may include but is not limited to 
\begin{itemize}
	\item[--] conformity assessment (see also JCGM~106 \cite{JCGM106})
	\item[--] propagation of uncertainty using the law of propagation of uncertainty \cite{GUM,GUM-S2}
	\item[--] propagation of uncertainty using the Monte Carlo method \cite{GUM-S1,GUM-S2}
\end{itemize}
For the conformity assessment, the coverage interval has been provided. If it is needed at a different probability level, or with a different distribution of the probability (e.g., one-sided), the estimate and standard uncertainty can be taken to obtain the parameters of the beta distribution (equations \eqref{eq:alpha} and \eqref{eq:beta}), followed by calculating the desired coverage interval (see for example listing~\ref{lst:code1} or listing~\ref{lst:code2}). 

In the case of using the law of propagation of uncertainty, the information on the certificate provides all that is needed without the need to make any computation: the estimate and the standard uncertainty.

Finally, when using the Monte Carlo method one needs to sample from the beta distribution with the appropriate values for the coefficients. If these are not provided explicitly, these can be obtained from the estimate and standard uncertainty using equations \eqref{eq:alpha} and \eqref{eq:beta}.

\section{Concluding remarks}

The revised ISO~19229 \cite{ISO19229:2018} provides not only better guidance in conducting experiments for purity analysis in the support of calibration gas mixture preparation and use in the calibration of analysers, but also a clause on how to calculate coverage intervals. In the vicinity of zero, it is appropriate to approximate the probability density function of the amount-of-substance fraction of an impurity by the beta distribution, as for small relative standard uncertainties (below, say, \SI{10}{\percent}) the shape of this probability distribution is very similar to the normal distribution and hence produces very similar \SI{95}{\percent} coverage intervals. As there is no direct relationship between the bounds of the non-symmetric coverage interval and the standard uncertainty, the latter should be reported explicitly to facilitate users of the law of propagation of uncertainty of the GUM in a subsequent uncertainty calculation. 

The use of the beta distribution has raised some issues as to whether it models properly the probability distribution of an amount-of-substance fraction. Deliberately, this question has been answered by benchmarking the beta distribution against the otherwise used normal distribution. This approach greatly simplified answering the question, for the appropriateness of the normal distribution was put aside. That proof would have been much harder to deliver. A nice feature of the beta distribution is that it approximates the normal distribution quite closely, so that the decision as to whether a result is in the vicinity of zero is not that critical. In case of doubt, the beta distribution should be used. 

\section*{Acknowledgement}

The work presented in this paper was funded by the Ministry of Economic Affairs and Agriculture of the Netherlands. 

\appendix

\section{Code used in calculations}

\subsection{Coverage intervals} \label{sec:CI}

The code in listing~\ref{lst:code1} shows how to calculate the coverage intervals based on the normal distribution and beta distribution in \texttt{R} \cite{R2017}. Lines 1-2 declare the value and standard uncertainty of the amount-of-substance fraction. Lines 4-5 compute the parameters $ \alpha $ and $ \beta $ of the beta distribution. Line 9 calculates the coverage interval using the normal distribution, taking as parameters the value and standard uncertainty of the amount-of-substance fraction. Line 10 calculates the coverage interval using the beta distribution taking $ \alpha $ and $ \beta $ as parameters. 

\begin{lstlisting}[language=R,style=interfaces,caption={Code for calculating probabilistic-symmetric coverage intervals for the normal and beta distributions},label={lst:code1}]
x.val = 300e-9
x.unc = 0.3*x.val

(alpha = ((1-x.val)/x.unc^2-1/x.val)*x.val^2)
(beta = alpha*(1/x.val-1))

prob = 0.99
# normal distribution 
(limits = qnorm(c((1-prob)/2,(1-prob)/2+prob),mean = x.val,sd = x.unc))
# beta distribution
(limits = qbeta(c((1-prob)/2,(1-prob)/2+prob),alpha,beta))
\end{lstlisting}

\subsection{Shortest coverage interval} \label{sec:ShortestCI}

Instead of using a trial-and-error approach, the shortest coverage interval can be found by using, for example Brent's method \cite{NR} implemented in the \texttt{R} function \texttt{optim()}. Lines 1-4 in listing~\ref{lst:code2} creates the function to be minimised, which computes the width of the coverage interval using the beta distribution for a given probability \texttt{prob} and lower limit of the coverage interval \texttt{L}. \texttt{alpha} and \texttt{beta} are the parameters of the beta distribution. The actual minimisation is done by a call to the \texttt{R} function \texttt{optim()} in line~6. Line~7 uses the location of the minimum (returned in \texttt{result\$par}) to compute the shortest coverage interval of the beta distribution.  

\begin{lstlisting}[language=R,style=interfaces,caption={Code for calculating the shortest coverage interval for the beta distribution},label={lst:code2}]
width.cf <- function(L,prob,alpha,beta) {
	limits = qbeta(c(L,L+prob),alpha,beta)
	limits[2]-limits[1]
}

(result = optim(par=0.025,fn = width.cf,prob=prob,alpha=alpha,beta=beta,method = "Brent",lower=0,upper=1-prob))
(limits = qbeta(c(result$par,result$par+prob),alpha,beta))
\end{lstlisting}


\end{document}